\documentclass[nohyper,twoside]{JHEP3}

\usepackage{graphicx}
\usepackage{amsmath}
\newcommand{\be}{\begin{equation}}
\newcommand{\ee}{\end{equation}}
\newcommand{\Dsla}{\ensuremath \raisebox{0.025cm}{\slash}\hspace{-0.32cm} D}
\newcommand{\reff}[1]{(\ref{#1})}

\title{Parity realization in Vector-like theories from Fermion Bilinears}

\author{V.~Azcoiti \\ 
Departamento de F\'{\i}sica Te\'orica, Universidad de
Zaragoza, Cl. Pedro Cerbuna 12, E-50009 Zaragoza (Spain) \\
E-mail: \email{azcoiti@azcoiti.unizar.es}} 
\author{G.~Di~Carlo \\ 
INFN, Laboratori Nazionali del Gran Sasso, 
67010 Assergi,(L'Aquila) (Italy) \\
E-mail: \email{gdicarlo@lngs.infn.it}} 
\author{A.~Vaquero \\
Departamento de F\'{\i}sica Te\'orica, Universidad de
Zaragoza, Cl. Pedro Cerbuna 12, E-50009 Zaragoza (Spain) \\
E-mail: \email{alexv@unizar.es}}

\abstract{We reconsider in this paper the old aim of trying to understand if 
the observed realization of discrete symmetries as Parity or $CP$ in the 
$QCD$ vacuum can be satisfied from first principles. We show how under the 
appropiate assumptions implicitely done by Vafa and Witten in their old 
paper on parity realization in vector-like theories, all parity and 
$CP$ odd operators constructed from fermion bilinears of the form 
$\bar\psi\tilde O\psi$ should take a vanishing vacuum expectation value 
in a vector-like theory with $N$ degenerate flavours ($N>1$). In our analysis 
the Vafa-Witten theorem on the impossibility to break spontaneously the 
flavour symmetry in a vector-like theory plays a fundamental 
role.}

\begin{document}

\section{Introduction}

The realization of symmetries in Nature is an essential point to 
understand all the phenomenology of elementary particles. Whereas 
space-time rotational invariance seems to be always realized, this is
not the case for the discrete parity, $P$, the charge conjugation, $C$,
and the time reversal, $T$, symmetries.
$P$ and $C$ symmetries are violated by electroweak processes and the only
fundamental theoretical restriction comes from the $CPT$ theorem,
which forbids a violation of the $CPT$ symmetry in relativistic 
field theory, even by the vacuum or ground state.

On the other hand the large amount of experimental results on the strong
interaction processes strongly suggests that $P$ and $C$ are preserved by
this interaction, vacuum included. The extremely small experimental bound 
on the neutron dipolar electric moment suggests that a $CP$ violating 
topological term should be excluded from the $QCD$ Lagrangian; and this
result, combined with the solution of the $U(1)$ problem via the axial
anomaly has generated the well known strong $CP$ problem \cite{scp}.
Hence the understanding of the realization of these symmetries in QCD is 
one of the challenges of Theory physicists.

One important ingredient of the fermion gauge interaction in $QCD$, as
opposite to the electroweak interaction, is the fact that it is vector-like.
Years ago Vafa and Witten \cite{vw1} {\it conjectured} that parity should not
be spontaneously broken in any vector-like theory as, for instance, $QCD$.

Althought the Vafa and Witten result has been taken for a long time by the
scientific community as a theorem, we call it conjecture for the following
two reasons:

{\it i.} V.A. and A. Galante have shown in \cite{va} that in the Vafa-Witten
work there is an implicit assumption: the Euclidean QCD Lagrangian 
in presence of an external source coupled to any $P=-1$ local pure gluonic 
operator is a  well defined quantum system, or at least a well defined
statistical system, but, as a rule, this needn't be true and the system may not be 
well defined. This assumption may seem innocuous, as suggested 
in \cite{cohen}, or even in \cite{ew}, where the authors argued that these ill-defined 
theories are not physical, but we would like to stress that at least for a very simple 
statistical system, the two-dimensional Ising model within an imaginary external 
magnetic field, the theorem does not hold. Indeed as it was shown in \cite{va}, 
if parity were spontaneously broken in QCD, the theory in presence of an external 
local $P=-1$ pure gluonic operator would be ill-defined. Similar concerns on the validity 
of the Vafa-Witten theorem were raised in \cite{xji}, and although in \cite{cr} some 
arguments against \cite{va} were exposed, the author claimed to have disproved the 
controversial theorem.

{\it ii.} There are explicit counter-examples for operators built from 
fermion fields, as the parity-flavour breaking term involved in the
Aoki phase \cite{aoki} for Wilson fermion lattice $QCD$. Even though in the 
Vafa-Witten
original paper it is stated in a foot note that the extension to fermionic
operators, after the integration of fermionic degrees
of freedom, is straightforward; this is not true: 
their argument can not be extended to order 
parameters constructed from fermion fields. This point, which has been 
stressed in \cite{sharpe}, will also be discussed in this paper.

The aim of this paper is to clarify a little bit these points and,
mainly, to extend the Vafa-Witten conjecture to fermion bilinear order
parameters, under suitable hypothesis.

In section 2 we briefly review the Vafa-Witten argumentation, and give an 
alternative way to reproduce their result, with the help of the probability
distribution function (hereafter $p.d.f.$) of the order parameter.
We discuss also in this section why the Vafa-Witten result can not be
extended in a simple way to order parameters constructed from fermion
bilinears. Since our extension of Vafa-Witten result to fermion
bilinears is based in the $p.d.f.$ of these operators, we give also a short 
review, in section 3, of the formalism developed in \cite{pdf} to define
the $p.d.f.$ of fermionic operators. In section 4 we apply this formalism
to the extension of the Vafa-Witten result to fermion bilinear order 
parameters, always under the assumption that a quantum theory can be 
consistently constructed in presence of these operators in the theory 
lagrangian. Section 5 contains our conclusions.

\section{The Vafa-Witten result: an alternative way}

The Euclidean partition function of a vector-like gauge theory can be
written as the following path integral over the Grassmann and gauge degrees
of freedom,

\be
{\cal Z} =  \int [dA_\mu^a ][d \bar \psi d \psi]  \exp \{-S_{YM}
+ \bar\psi ( \Dsla+m) \psi \}  \label{partf}
\ee

\noindent
where $S_{YM}$ is the action for the Yang-Mills fields and the gauge-fermion
coupling is a bilinear of the fermion fields. $\Dsla$ is the Euclidean Dirac
operator and what characterizes vector-like theories is that the fermion
determinant $\det(\Dsla+m)$, which appears in the effective gauge theory
after integration of the fermion fields, is positive-definite for any non
vanishing bare fermion mass.

If we add now to the action a hermitian $P=-1$ local order parameter 
$Y(A_\mu^a)$, which depends only on the gauge fields, as for instance a 
topological $\theta$-vacuum term, the Euclidean partition function for
the effective gauge theory is:

\be
{\cal Z} =  \int [dA_\mu^a ] \exp \{-S_{YM} -\lambda \int d^4x
\>\>  Y(x) +\ln \det(\Dsla+m) \}  \label{pfop}
\ee

Years ago Vafa and Witten \cite{vw1} gave an argument against spontaneous 
$P$ breaking in vector-like parity-conserving theories, as $QCD$.
The main point in their argument was the observation that any arbitrary 
hermitian local order parameter, $Y(A_\mu^a)$, constructed only from Bose
fields, should be proportional to an odd power of the four indices 
totally antisymmetric tensor $\epsilon^{\mu\nu\rho\sigma}$ and therefore
it would pick up a factor of $i$ under Wick rotation. The local order
parameter $Y(A_\mu^a)$ can therefore be written as $Y(A_\mu^a)=i X(A_\mu^a)$
with $X$ real, and the Euclidean partition function \reff{pfop} becomes:

\be
{\cal Z} =  \int [dA_\mu^a ] \exp \{-S_{YM} - i \lambda \int d^4x
\>\>  X(x) +\ln \det(\Dsla+m) \}  \label{pfop2}
\ee

The addition of an external symmetry breaking field $\lambda Y(A_\mu^a)$
to the Lagrangian in Minkowski space  becomes then a pure phase factor
in the path integral definition of the partition function in Euclidean 
space \reff{pfop2}. But a pure phase factor in the integrand of a partition 
function with positive-definite integration measure can only increase the
vacuum energy density (free energy density). Vafa and Witten concluded that, 
in such a situation, the mean value of the order parameter should vanish
in the limit of vanishing symmetry breaking field.

The only objection done in \cite{va} was that any argumentation
on the vacuum energy density (free energy density) of the theory in presence 
of a parity breaking term $\lambda$ requires the previous assumption that
the quantum theory is well defined in such a condition. This statement is
less naive that what it might seem at first sight; indeed, as it was shown in
\cite{va}, the existence of a non-parity-invariant vacuum state would imply that the
theory in presence of a parity-breaking term is ill-defined. On the other hand, if we
take this assumption as a requirement, the conjecture becomes a full-fledged theorem,
which states that, if the theory with any added $P=-1$ external bosonic sources is 
well-defined, parity can not be spontaneously broken.

Let us see now an alternative way of getting the Vafa-Witten result, that 
makes use of the concept of the $p.d.f.$ of a local operator. We do not intend to give
an alternative proof in which we solve the issues arised in \cite{va}, but to show a
different path to reach the same result, introducing the $p.d.f.$ formalism. In our derivation
we will assume that we can consistently define the quantum theory with a $P$-breaking local
order parameter term.

Let be $Y(A_\mu^a)$ a local operator constructed with Bose fields. The
probability distribution function of this local operator in the effective
gauge theory described by the partition function \reff{pfop} is

\be
P(c)=\left\langle \delta\left(c- {1 \over V} \int d^4x \>\> Y(x) \right) \right\rangle 
\label{pdic}
\ee

\noindent
where $V$ is the space-time volume and the mean values are computed over
all the Yang-Mills configurations using the integration measure defined 
in \reff{pfop}. Equivalently one can define its Fourier transform

\be
P(q)=\int e^{iqc} P(c) dc
\label{pdiq}
\ee

\noindent
which, in our case, is given by the following expression

\be
P(q)= {{\int [dA_\mu^a ] \exp \{-S_{YM} +({iq \over V} -\lambda) \int d^4x
\>\>  Y(x) +\ln \det(\Dsla+m) \}} \over 
{\int [dA_\mu^a ] \exp \{-S_{YM} -\lambda \int d^4x
\>\>  Y(x) +\ln \det(\Dsla+m) \}} }
\label{pdiqex}
\ee

\noindent
or, in a short notation:

\be
P(q)= \left\langle \exp ({iq \over V} \int d^4x \>\> Y(x) ) \right\rangle
\label{pdiqsh}
\ee

If we calculate $P(c)$ for a local $P$-breaking operator in
absence of a $P$-breaking term in the action of \reff{pfop}
$(\lambda=0)$ we expect a Dirac $\delta$ distribution centered
at the origin if the vacuum state is non degenerate. On the
contrary, if the vacuum state is degenerate. the expected form for $P(c)$
is:

\be
P(c)=\sum_\alpha w_\alpha \> \delta (c-c_\alpha)
\label{degen}
\ee

\noindent
where $c_\alpha$ is the mean value of the local order parameter in the 
vacuum state $\alpha$ and $w_\alpha$ are positive real numbers which give
us the probability of each vacuum state ($\sum_\alpha w_\alpha =1 $).

If the only reason to have a degenerate vacuum is the spontaneous
breaking of a discrete $Z_2$ symmetry, we expect in the more standard
case two symmetric vacuum states ($\pm c_\alpha$) with the same weights
$w_\alpha$ ( in the most general case an even number of vacuum states with
opposite values of $c_\alpha$ ).  The probability distribution function
$P(c)$ will be then the sum of two symmetric Dirac $\delta$ with equal
weights:

\be
P(c)={1 \over 2} \delta (c-c_\alpha) + {1 \over 2} \delta (c+c_\alpha)
\label{degen2}
\ee

\noindent
and its Fourier transform

\be
P(q)=\cos(q c_\alpha)
\label{qdeg2}
\ee

Now we are ready to apply this formalism to the case we are interested in.
The relevant fact now, as in the Vafa-Witten paper, is the fact
that the local $P$-breaking order parameter is a pure imaginary number
$Y(A_\mu^a)=i X(A_\mu^a)$). In such a case we get for $P(q)$:

\be
P(q)= \left\langle \exp (-{q \over V} \int d^4x \>\> X(x) ) \right\rangle
\label{pdiqshx}
\ee

\noindent
but the mean value of a real and positive quantity
computed with a well defined probability distribution function (positive
integration measure) can not be negative. Were the symmetry spontaneously
broken we should get for $P(q)$ a cosine function \reff{qdeg2} (a
sum of cosines in the most general case) which takes positive and negative
values. Since this case is excluded, $P(q)$ should be a constant function 
equal to $1$, representing a symmetric vacuum state.

The generalization of the Vafa-Witten argumentation to  fermionic
bilinears upon integrating out the fermion fields, as suggested in
\cite{vw1}, can not be carried out, as discussed in \cite{sharpe}.
Indeed if we add a fermionic bilinear $P=-1$ term, $\lambda i \bar\psi
\gamma_5 \psi$, to the fermionic action, we get in the effective Yang-Mills 
theory an extra-term to the pure gauge action with a non-vanishing
$\lambda$-dependent real part. Thus the contribution to the effective
gauge theory of the symmetry breaking term in the integrand of the
partition function is no longer a pure phase factor and therefore
how this term modifies the vacuum energy-density cannot be stated 
{\it a priori}. In fact there is a well known case, the Aoki phase \cite{aoki}
in lattice QCD with two degenerate flavours of Wilson fermions, in which 
notwithstanding the integration measure is non negative, Parity is 
spontaneously broken.

\section{ The $p.d.f.$ for fermionic local operators}

Our aim is to extend, as much as possible, the Vafa-Witten result 
for pure gluonic operators to fermion bilinear local operators.
To this end we will make use, as in the previous section, of the
$p.d.f.$ of local operators constructed with Grassmann fields.
Since this formalism is not, in our opinion, very standard, we want to
devote this section to summarize the main results reported in \cite{pdf}
on this subject.

The motivation to develop this formalism was precisely the study of
the vacuum invariance (non invariance), in quantum theories regularized on a 
space time lattice, under symmetry transformations which, as chiral, flavour 
or baryon symmetries, involve fermion fields. The numerical determination of
the $p.d.f.$ of the order parameter is a standard procedure when analyzing 
spontaneous symmetry breaking in spin systems or in quantum field theories
with Bose fields. Indeed these degrees of freedom can be
simulated directly on a computer. However in the numerical simulation
of a Lattice Gauge Theory (LGT) with dynamical fermions the Grassmann
fields, which at present can not be simulated in a computer, must be
integrated analytically. Then even if the ground state of the theory is 
not invariant under the chiral, flavour or baryon symmetries, in the
analytical procedure we integrate over all possible vacuum states, 
the order parameter vanishing always independently of the symmetry realization.
This is the reason why the standard procedure to study, for instance, the 
chiral properties of the $QCD$ vacuum in numerical simulations on a lattice 
is to add a mass term to the Euclidean action, which breaks chiral symmetry
and to compute the chiral condensate $\langle\bar\psi\psi\rangle$ as a
function of the mass $m$ performing, at the end, a numerical extrapolation
to the chiral limit in order to see if $\langle\bar\psi\psi\rangle$ takes a 
non-vanishing value in this limit.

Notwithstanding that Grassmann variables cannot be simulated in a
computer, it was shown in \cite{pdf} that an analysis of spontaneous
symmetry breaking based on the use of the $p.d.f.$ of fermion local
operators, and therefore free from extrapolations in the symmetry breaking 
field, can also be done in $QFT$ with fermion degrees of freedom.

The procedure is similar to the one used in the previous section for
bosonic degrees of freedom. The starting point is to choose an order 
parameter for the desired symmetry $O(\psi,\bar\psi)$ (typically a
fermion bilinear) and characterize each vacuum state $\alpha$ by the
expectation value $c_\alpha$ of the order parameter in the $\alpha$ state

\be
c_\alpha= {1 \over V} \int \langle O(x)\rangle_\alpha d^4 x
\label{exval}
\ee

The $p.d.f.$ $P(c)$ of the order parameter will be given by

\be
P(c)=\sum_\alpha w_\alpha \> \delta (c-c_\alpha)
\label{pdcferm}
\ee

\noindent
which can also be written as \cite{pdf}

\be
P(c)=\left\langle \delta({1 \over V} \int O(x) d^4x -c) \right\rangle
\label{pdcferm2}
\ee

\noindent
the mean value computed with the integration measure of the path
integral formulation of the Quantum Theory.

The Fourier transform $P(q)=\int e^{iqc} P(c) dc$ can be written, for the 
theory described by partition function \reff{partf}, as

\be
P(q)= {{\int [dA_\mu^a ] [d\bar\psi d\psi]
\exp \{-S_{YM} +\bar\psi (\Dsla+m)\psi+ {iq \over V} \int d^4x
\>\>  O(x)\}} \over {\int [dA_\mu^a ] [d\bar\psi d\psi] \exp 
\{-S_{YM}  +\bar\psi(\Dsla+m)\psi \}} }
\label{pdiqferm}
\ee

In the particular case in which $O$ is a fermion bilinear of $\bar\psi$
and $\psi$

\be
O(x)=\bar\psi(x) \tilde O \psi(x)
\label{bili}
\ee

\noindent
with $\tilde O$ any matrix with Dirac, color and flavour indices, equation
\reff{pdiqferm} becomes

\be
P(q)= {{\int [dA_\mu^a ] [d\bar\psi d\psi]
\exp \{-S_{YM} +\bar\psi (\Dsla+m+{iq \over V} \tilde O)\psi \}} 
\over {\int [dA_\mu^a ] [d\bar\psi d\psi] \exp 
\{-S_{YM}  +\bar\psi(\Dsla+m)\psi \}} }
\label{pdiqbili}
\ee

Integrating out the fermion fields in \reff{pdiqbili} one gets

\be
P(q)= {{\int [dA_\mu^a ] e^{-S_{YM}} \det(\Dsla+m+{iq \over V} \tilde O) } 
\over {\int [dA_\mu^a ] e^{-S_{YM}} \det(\Dsla+m) } }
\label{pdiqbi}
\ee

\noindent
which can expressed also as the following mean value

\be
P(q)=\left\langle {{\det (\Dsla+m+{iq \over V} \tilde O)} \over
{\det(\Dsla+m)}} \right\rangle
\label{pdiqb}
\ee

\noindent
computed in the effective gauge theory with the integration measure

$$
[dA_\mu^q] e^{-S_{YM}} \det(\Dsla+m)
$$

The particular form expected for the $p.d.f.$ $P(c)$, $P(q)$, 
depends on the realization of the corresponding symmetry in the
vacuum. A symmetric vacuum will give

\be
P(c)=\delta(c)
\label{delta}
\ee
$$
P(q)=1
$$

\noindent
whereas, if we have for instance a $U(1)$ symmetry which is
spontaneously broken, the expected values for $P(c)$ and $P(q)$
are \cite{pdf}

$$
P(c)=[ \pi (c_0^2-c^2)^{1/2} ]^{-1} \>\>\>\>  -c_0< c< c_0
$$
$$
P(c)=0  \>\>\>\>  {\rm otherwise}
$$
\be
P(q)={1 \over 2\pi} \int_{-\pi}^\pi d\theta e^{iqc_0\cos\theta}
\label{pdfu1}
\ee

\noindent
the last being the well known zeroth order Bessel function of the
first kind, $J_0(qc_0)$.

In the simpler case in which a discrete $Z(2)$ symmetry, as Parity, is 
spontaneously broken, the expected form is

$$
P(c)= {1 \over 2} \delta(c-c_0) + {1 \over 2} \delta(c+c_0) 
$$
\be
P(q)=\cos(q c_0)
\label{pdfz2}
\ee

\noindent
or a sum of symmetric delta functions ($P(c)$) and a sum of
cosines ($P(q)$) if there is an extra vacuum degenerancy.

This formalism has been successfully applied to the analysis of spontaneous 
chiral symmetry breaking in LGT with Kogut-Susskind fermions \cite{pdf}, as
well as to find a new phase in finite baryon density two-colors QCD with 
diquark condensation and spontaneous breaking of baryon symmetry \cite{dq}.

The analysis of the diquark condensation phase in finite density 
two-color QCD with four Kogut-Susskind flavours done in \cite{dq}
showed also one of the potentialities of this approach.
Indeed the order parameter used in the search for diquark condensation phase

$$
\psi\tau_2\psi+\bar\psi\tau_2\bar\psi
$$

\noindent
when added to the $SU(2)$ Lagrangian as an external source that breaks
baryon symmetry, place us against a tedious problem, the well known
sign problem. After integrating out the fermion fields we get the 
Pfaffian of a matrix and this Pfaffian is not positive definite. The
standard approach to perform numerical analysis of spontaneous symmetry
breaking, based on the computation of the condensate at non-vanishing
external source, does not work because standard Monte Carlo techniques
apply to systems with a well defined (positive definite) Boltzmann
factor. The $p.d.f.$ of the diquark condensate is computed, on the contrary, 
at vanishing external source, and the sign problem is absent in this case.

We are ready now to apply this approach in the next section to the
analysis of spontaneous $P$ breaking in vector-like theories, by investigating
order parameters constructed from bilinears of the fermion field.

\section{Vacuum expectation values of $P=-1$ fermion bilinears}

In this section we will show that all local bilinear $P=-1$ 
gauge invariant operators 
of the form $\bar\psi\tilde O\psi$, with $\tilde O$ a constant matrix with
Dirac, color and flavour indices, should take a vanishing vacuum 
expectation value in any vector-like theory with N degenerate flavours.

Let us start with the one-flavour case since, as it will be shown, it is a 
special case. The standard hermitian local and gauge invariant $P$ order 
parameter bilinear in the fermion fields is:

$$
\bar\psi\tilde O\psi= i\bar\psi\gamma_5\psi
$$

Now we can apply the result of the previous section, and in particular
equation \reff{pdiqb}, which give us the $p.d.f$ of $\tilde O$ in momentum
space

\be
P(q)=\left\langle {{\det (\Dsla+m+{q \over V} \gamma_5)} \over
{\det(\Dsla+m)}} \right\rangle
\label{pdiq1f}
\ee

The determinant of the Dirac operator in the denominator of \reff{pdiq1f}
is positive definite, but the numerator of this expression, even if real,
has not well defined sign.
The final form for $P(q)$ will depend crucially on the distribution of
the real eigenvalues of $\gamma_5(\Dsla+m)$. Therefore we cannot say
{\it a priori} whether $P(q)$ will be the constant function $P(q)=1$ 
(symmetric vacuum) or a cosine function (spontaneously broken $P$).

Let us go now to the $N$ flavour case ($N>1$). The most general
$P=-1$ hermitian and gauge invariant local order parameters 
$\bar\psi\tilde O\psi$ which can be constructed are:

$$
i\bar\psi\gamma_5\psi
$$
\be
i\bar\psi\gamma_5\bar\tau\psi
\label{oti2f}
\ee

\noindent
with $\bar\tau$ any of the hermitian generators of the $SU(N)$ flavour
group. However, since flavour symmetry cannot be spontaneously broken
in a vector-like theory \cite{vw2}, we will restrict our analysis to 
the flavour singlet case.

\be
i\bar\psi\gamma_5\psi = i\bar\psi_u\gamma_5\psi_u + i\bar\psi_d\gamma_5\psi_d
+ i\bar\psi_s\gamma_5\psi_s + ...
\label{otising}
\ee

Let us assume that $\langle  i\bar\psi_u\gamma_5\psi_u \rangle =\pm c_0
\neq 0$. Since flavour symmetry is not spontaneously broken we should have:

\be
\langle i\bar\psi_u\gamma_5\psi_u \rangle = 
\langle i\bar\psi_d\gamma_5\psi_d \rangle
= \langle i\bar\psi_s\gamma_5\psi_s \rangle = ...
\label{flavsim}
\ee

Thus the system will show two degenerate vacua with all the condensates
oriented in the up (down) direction.

We want to remark that the ``ferromagnetic'' nature of these two
vacua is imposed by the realization of flavour symmetry in the vacuum.
Otherwise one could imagine also ``antiferromagnetic'' vacua with
antiparallel condensates, or even more complex structures.

In vector-like gauge theories the interaction between different flavour
is mediated by gluons. Then, assuming 
$\langle i\bar\psi_u\gamma_5\psi_u\rangle \neq 0$ the impossibility to break
flavour symmetry \cite{vw2} could suggest that the gauge interaction would
favor ``ferromagnetic'' vacua with parallel oriented condensates.
However the actual dynamics can also be more complex. Indeed we will argue
now that a non vanishing condensate $\langle i\bar\psi_u\gamma_5\psi_u\rangle
 \neq 0$, which would imply spontaneous breaking of $P$, $CP$, $T$ and $CT$,
can be excluded.

Let us apply equation \reff{pdiqb} to the computation of the $p.d.f.$ 
in momentum space $P_{ud}(q)$ of $i\bar\psi_u\gamma_5\psi_u + 
i\bar\psi_d\gamma_5\psi_d$.

\be
P_{ud}(q)=\left\langle \left({{\det (\Dsla+m-{q \over V} \gamma_5)} \over
{\det(\Dsla+m)}}\right)^2 \right\rangle
\label{pdiq2f}
\ee

\noindent
where $\Dsla+m$ in \reff{pdiq2f} is the one flavour Dirac operator and the
mean value is computed in the theory with N-degenerate flavours.

Equation \reff{pdiq2f} give us the mean value, computed with a positive 
definite integration measure, of a real non-negative quantity. Thus 
$P_{ud}(q)$ will be a positive definite, or at least a non negative definite,
quantity. But were $\langle i\bar\psi_u\gamma_5\psi_u\rangle \neq 0$ we
should expect a cosine function for $P_{ud}(q)$ since 
$i\bar\psi_u\gamma_5\psi_u$
and $i\bar\psi_d\gamma_5\psi_d$ are enforced to take the same $v.e.v.$ for
flavour symmetry. Since the positivity of $P_{ud}(q)$ excludes such a 
possibility, we can conclude that all the pseudo-scalar condensates 
$i\bar\psi_f\gamma_5\psi_f$ take vanishing expectation values.

In order to give a simple illustration of some of this ideas, one can think of 
a physical system composed by two replicas of the two-dimensional Ising model. 
The spins $s^a$ of replica $a$ and $s^b$ of replica $b$ live on the sites of a 
two-dimensional lattice. They are coupled inside each replica with the
standard nearest-neighbor ferromagnetic coupling, and we add an ultra-local
interaction between $s^a$ and $s^b$ at the same site.

The Hamiltonian of this system can be written in the following way:

\be
{\cal H} = -J \sum_{<i,j>} (s_i^as_j^a+s_i^bs_j^b) - k \sum_i s_i^as_i^b
\label{hisi2}
\ee

This system has the standard $Z(2)$ $P$ symmetry ( ${\cal H}$ is invariant
under the change of sign of all spins) plus an extra $Z(2)$ symmetry, the
replica symmetry, as a consequence of the invariance of ${\cal H}$ under 
the permutations of the $a$ and $b$ indices.

Let us neglect for a while the replica-replica interaction ($k=0$ in
\reff{hisi2}) and choose $F={J\over{KT}}>0.44$ (ferromagnetic phase). If we 
denote 
the magnetization per site of replica $a$ and $b$ as $m_a$ and $m_b$, we
can construct two order parameters: $\langle m_a-m_b\rangle$ and
$\langle m_a+m_b\rangle$. Both are $P=-1$ order parameters and the first one
is also order parameter for the $Z(2)$ replica symmetry.

Under the previous assumed conditions ($F>0.44$, $k=0$) we have four 
degenerate vacua ($++, --, +-, -+$) characterized by the following
values of the order parameters:

$$
\langle m_a+m_b\rangle_{++}=2m_0 \>\>\>\>\> \langle m_a-m_b\rangle_{++}=0
$$
$$
\langle m_a+m_b\rangle_{--}=-2m_0 \>\>\>\>\> \langle m_a-m_b\rangle_{--}=0
$$
$$
\langle m_a+m_b\rangle_{+-}=0 \>\>\>\>\> \langle m_a-m_b\rangle_{+-}=2m_0
$$
$$
\langle m_a+m_b\rangle_{-+}=0 \>\>\>\>\> \langle m_a-m_b\rangle_{-+}=-2m_0
$$

If we switch on now the replica-replica interaction ($k\neq 0$) we expect that
the vacuum degeneracy will be reduced. Indeed $k>0$ values will enforce a 
parallel orientation of spins of different replicas, whereas $k<0$ will
favor an anti-parallel orientation. In the first case we will get two
``ferromagnetic'' degenerate vacua preserving the replica symmetry. In
the second case we get two ``antiferromagnetic'' degenerate vacua with
spontaneous parity and replica breaking.

What we can learn from this simple model is that a small replica-replica
interaction is enough to break some of the vacuum degeneracy. Then
``ferromagnetic'' or ``antiferromagnetic'' degenerate vacua are selected,
depending on the ferro-antiferromagnetic character of the replica-replica
interaction.

\section{Conclusions}

We have reconsidered in this paper the old aim of trying to understand if 
the observed realization of discrete symmetries as Parity or $CP$ in the 
$QCD$ vacuum can be satisfied from first principles. Although this aim 
is at the moment too ambitious, we have shown how under the appropiate 
plausible assumptions implicitely done in \cite{vw1}, all parity and 
$CP$ odd operators constructed from fermion bilinears of the form 
$\bar\psi\tilde O\psi$ should take a vanishing vacuum expectation value 
in a vector-like theory with $N$ degenerate flavours ($N>1$). In our analysis 
the Vafa-Witten theorem on the impossibility to break spontaneously the 
flavour symmetry in a vector-like theory \cite{vw2} plays a fundamental 
role. Indeed our result does not apply to lattice $QCD$ with two degenerate 
Wilson fermions, where a phase with spontaneous Parity breaking has been 
found \cite{aoki}. It follows because, notwithstanding the integration measure 
is non negative, the Vafa-Witten theorem on the impossibility of breaking 
spontaneously a vector symmetry in a vector-like theory is not 
applicable to this case, for the Wilson quark propagator is not 
bounded in an arbitrary background gauge field. Indeed flavour 
symmetry is spontaneously broken in the Aoki phase.

\acknowledgments

This work has been partially supported by an INFN-MEC collaboration 
and CICYT (grant FPA2006-02315).

\end{document}